\documentclass{article}

\usepackage{arxiv}

\usepackage[utf8]{inputenc} 
\usepackage[T1]{fontenc}    
\usepackage{hyperref}       
\usepackage{url}            
\usepackage{booktabs}       
\usepackage{amsfonts}       
\usepackage{amsmath}       
\usepackage{nicefrac}       
\usepackage{microtype}      
\usepackage{lipsum}
\usepackage{graphicx}

\title{A closed-form solution for the kinematics of asymmetric Miura vertices}

\author{
  Soroush Kamrava \\
  Mechanical and Industrial Engineering\\
  Northeastern University\\
  Boston, MA 02115 \\
  \texttt{kamrava.s@northeastern.edu} \\
    \And
 Chang Liu \\
  Mechanical and Industrial Engineering\\
  Northeastern University\\
  Boston, MA 02115 \\
  \texttt{liu.chang7@northeastern.edu} \\
    \And
 Alec Q.~Orlofsky \\
  Mechanical and Industrial Engineering\\
  Northeastern University\\
  Boston, MA 02115 \\
  \texttt{orlofsky.a@northeastern.edu} \\
   \And
 Ashkan Vaziri \\
  Mechanical and Industrial Engineering\\
  Northeastern University\\
  Boston, MA 02115 \\
  \texttt{vaziri@coe.neu.edu} \\
   \And
 Samuel M.~Felton \\
  Mechanical and Industrial Engineering\\
  Northeastern University\\
  Boston, MA 02115 \\
  \texttt{s.felton@northeastern.edu} \\
}

\begin{document}
\maketitle

\begin{abstract}
The Miura vertex is a versatile origami pattern found in a variety of mechanisms. Previous papers have derived and validated a closed-form solution for the kinematics of a symmetric Miura vertex, but the motion of an asymmetric vertex has only been shown numerically. In this paper, we present the trigonometric derivation of a closed-form solution for the folding of an asymmetric Miura vertex.
\end{abstract}

\keywords{Origami \and Miura \and Four Crease Vertex \and Closed-form Solution }

\section{Introduction}

The Miura fold is a canonical origami pattern \cite{miura2009science}. First developed and proposed as a method for packing a large solar array into a small volume, the pattern consists of a two-dimensional array of four-edge vertices. Although the unit vertex is traditionally symmetric, it can be asymmetric with two colinear creases (the `spinal' creases) at the flat configuration, see Figure 1. 

\begin{figure}[t]
	\includegraphics[width=140mm]{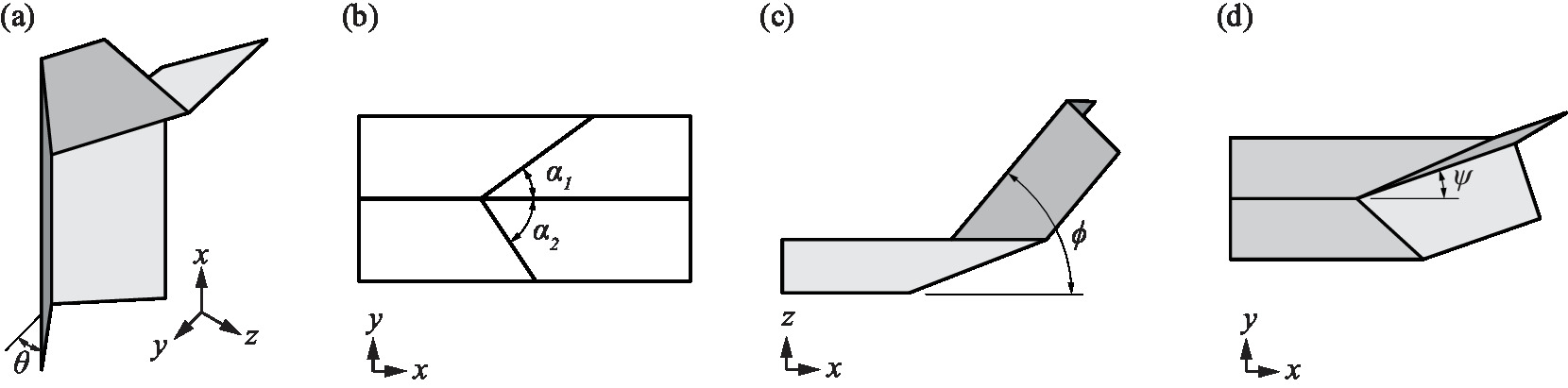}
	\centering
	\caption{(a) The asymmetric Miura vertex `bends' between its two spinal creases when one crease is folded by an angle $\theta$. (b) We can define the origami pattern by the angles $\alpha_1$ and $\alpha_2$ between the peripheral and spinal creases. (c) All Miura vertices deflect by an angle $\phi$ around the $y$-axis. (d) Asymmetric vertices also deflect by an angle $\psi$ around the $z$-axis.}
	\label{Figure1}
\end{figure}

In addition to solar arrays, Miura vertices have been studied in a variety of other contexts, taking advantage of their tunable kinematics and single degree of freedom to create complex and constrained mechanisms \cite{fang2017dynamics,zuliani2018minimally,liu2018transformation}. One family of mechanisms is the origami string, which is a one-dimensional array of Miura vertices that can approximate a path in 2D or 3D space \cite{kamrava2018programmable,kamrava2018slender}. This is accomplished by considering each spinal crease a line segment, and selecting the appropriate fold angle $\theta$ and pattern angles $\alpha_i$ at each vertex to achieve the desired angular displacement (given as $\phi$ and $\psi$) between adjacent spinal creases.

A closed-form solution for the angular displacements ($\phi$ and $\psi$) as a function of the fold pattern ($\alpha_1$ and $\alpha_2$) and folding angle ($\theta$) is valuable when designing Miura strings and other Miura-based mechanisms. Previous work has derived a closed-form solution for symmetric vertices \cite{miyashita2015multi,kamrava2017origami}. Here, we derive a closed form solution for the more general case of asymmetric vertices.

\section{Geometric Derivation}

\begin{figure}[b]
	\includegraphics[width=65mm]{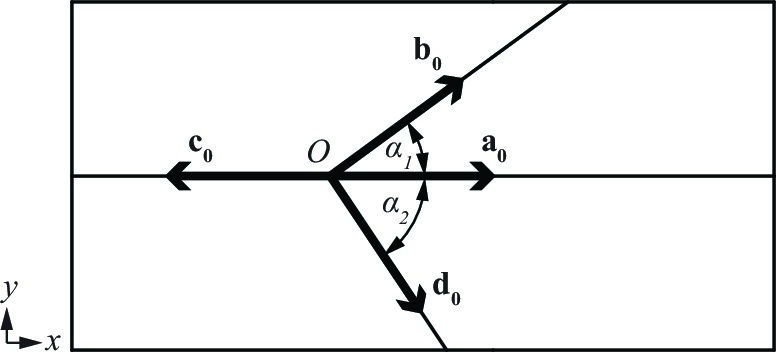}
	\centering
	\caption{In this derivation each vertex is represented by a unit vector.}
	\label{Figure2}
\end{figure}

We first represent each crease of the vertex as a vector (Fig.~\ref{Figure2}). Consider four unit vectors that are colinear with the four creases.
\begin{align}
\hat{\mathbf{a}} &= \left( a_x, a_y, a_z \right) \\
\hat{\mathbf{b}} &= \left( b_x, b_y, b_z \right) \\
\hat{\mathbf{c}} &= \left( c_x, c_y, c_z \right) \\
\hat{\mathbf{d}} &= \left( d_x, d_y, d_z \right)
\end{align}

Without loss of generality, we can fix the vertex at the origin $(0,0,0)$ and fix $\hat{\mathbf{c}}$ collinear with the X-axis.

\begin{figure}[t]
	\includegraphics[width=70mm]{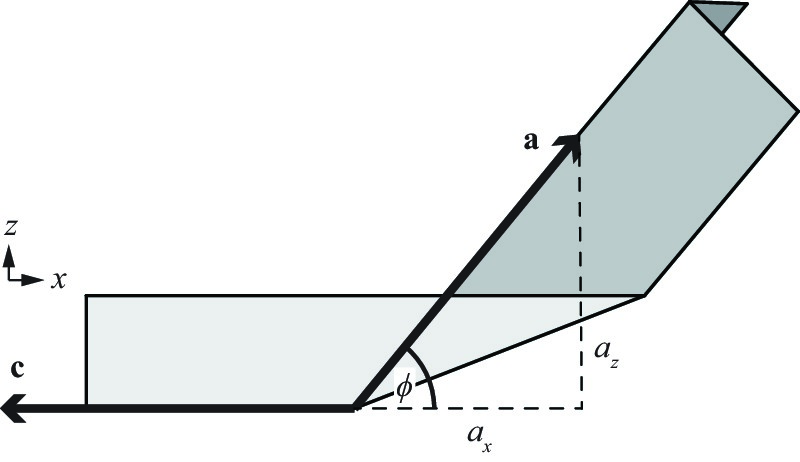}
	\centering
	\caption{We solve for $\phi$ by finding the vector components $a_x$ and $a_z$}
	\label{Figure3}
\end{figure}

\begin{figure}[b]
	\includegraphics[width=70mm]{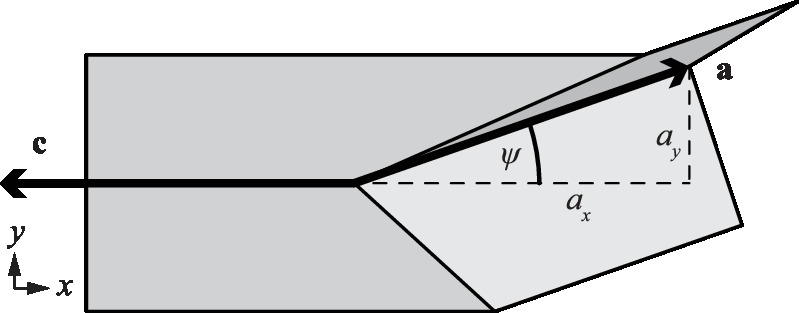}
	\centering
	\caption{We solve for $\psi$ by finding the vector components $a_x$ and $a_y$}
	\label{Figure4}
\end{figure}

When the vertex is flat in the X-Y plane

\begin{align}
\hat{\mathbf{a}}_0 &= \left( 1, 0 \right) = \langle 1, \angle 0 \rangle \\
\hat{\mathbf{b}}_0 &= \left( \cos \alpha_1, \sin \alpha_1 \right) = \langle 1, \angle \alpha_1 \rangle \\
\hat{\mathbf{c}}_0 &= \left( -1, 0 \right) = \langle 1, \angle \pi \rangle \\
\hat{\mathbf{d}}_0 &= \left( \cos \alpha_2, -\sin \alpha_2 \right) = \langle 1, \angle -\alpha_2 \rangle
\end{align}

When folded, the vectors $\hat{\mathbf{b}}$ and $\hat{\mathbf{d}}$ rotate around the X-axis by angles $\theta$ and $-\theta$, respectively.

\begin{align}
    \hat{\mathbf{b}} &= \left( \cos \alpha_1, \sin \alpha_1 \cos \theta, \sin \alpha_1 \sin \theta \right) \\
    \hat{\mathbf{d}} &= \left( \cos \alpha_2, -\sin \alpha_2 \cos \theta, \sin \alpha_2 \sin \theta \right)
\end{align}

The crease pattern angle $\alpha_1$ will always be the angle between vectors $\hat{\mathbf{a}}$ and $\hat{\mathbf{b}}$. Similarly, the crease pattern angle $\alpha_2$ will always be the angle between vectors $\hat{\mathbf{a}}$ and $\hat{\mathbf{d}}$.

\begin{align}
    \cos \alpha_1 &= \hat{\mathbf{a}} \cdot \hat{\mathbf{b}} = a_x \cos \alpha_1 + a_y \sin \alpha_1 \cos \theta + a_z \sin \alpha_1 \sin \theta \\
    \left( 1 - a_x \right) \cos \alpha_1 &= \sin \alpha_1 \left( a_y \cos \theta + a_z \sin \theta \right) \\
    \frac{1 - a_x}{\tan \alpha_1} &= a_y \cos \theta + a_z \sin \theta
\end{align}

\begin{align}
    \cos \alpha_2 &= \hat{\mathbf{a}} \cdot \hat{\mathbf{d}} = a_x \cos \alpha_2 - a_y \sin \alpha_2 \cos \theta + a_z \sin \alpha_2 \sin \theta \\
    \left( 1 - a_x \right) \cos \alpha_2 &= \sin \alpha_2 \left( - a_y \cos \theta + a_z \sin \theta \right) \\
    \frac{1 - a_x}{\tan \alpha_2} &= -a_y \cos \theta + a_z \sin \theta
\end{align}

We can combine these equations to solve for $a_y$ and $a_z$ in terms of $a_x$. Here we used dummy variables $K_1$ and $K_2$ to condense and simplify the equations.

\begin{align}
a_z \sin \theta &= \frac{1 - a_x}{2} \left( \frac{1}{\tan \alpha_1} + \frac{1}{\tan \alpha_2} \right) = \left(1 - a_x\right) K_1 \\
a_y \cos \theta &= \frac{1 - a_x}{2} \left( \frac{1}{\tan \alpha_1} - \frac{1}{\tan \alpha_2} \right) = \left(1 - a_x\right) K_2 \\
a_y &= \frac{K_2}{\cos \theta} \left( 1 - a_x \right) \\ 
a_z &= \frac{K_1}{\sin \theta} \left( 1 - a_x \right)
\end{align}

Because $\hat{\mathbf{a}}$ is a unit vector we can solve for $a_x$ explicitly. Here we use another dummy variable $K_3$ to condense and simplify the equations.

\begin{align}
a_x &= \sqrt{1 - a_y^2 - a_z^2} \\
a_x^2 &= 1 - \left( \frac{K_2}{\cos \theta} \left( 1 - a_x \right) \right)^2 - \left( \frac{K_1}{\sin \theta} \left( 1 - a_x \right) \right)^2 \\
&= 1 - \left[ \left( \frac{K_2}{\cos \theta} \right)^2 + \left( \frac{K_1}{\sin \theta} \right)^2  \right] \left( 1 - 2a_x + a_x^2 \right) = 1 - K_3 \left( 1 - 2a_x + a_x^2 \right) \\
\textrm{where} \qquad K_3 &= \left( \frac{K_2}{\cos \theta} \right)^2 + \left( \frac{K_1}{\sin \theta} \right)^2 \\
0 &= \left( K_3 + 1  \right) a_x^2 - 2 K_3 a_x + \left( K_3 - 1  \right) \\
a_x &= \frac{2K_3 \pm \sqrt{\left( 2 K_3 \right)^2 - 4\left( K_3 + 1 \right) \left( K_3 - 1 \right) }}{2\left( K_3 + 1 \right)} = \frac{K_3 \pm 1}{K_3 + 1}
\end{align}

The two solutions for $a_x$ represent the two possible configurations for the vertex. In one configuration, the peripheral creases (represented by $\hat{\mathbf{b}}$ and $\hat{\mathbf{d}}$) don't fold and the spinal creases stay collinear so that $\phi=0$. This corresponds with the following value for $a_x$:
\begin{align}
a_x &= \frac{K_3 + 1}{K_3 + 1} = 1
\end{align}

We are interested in the configuration where $\phi\neq 0$, so the remaining solution considers the alternate case:
\begin{align}
a_x &= \frac{K_3 - 1}{K_3 + 1} = \frac{\left( \frac{K_2}{\cos \theta} \right)^2 + \left( \frac{K_1}{\sin \theta} \right)^2  - 1}{\left( \frac{K_2}{\cos \theta} \right)^2 + \left( \frac{K_1}{\sin \theta} \right)^2 + 1}
\end{align}

We can use this identity to solve for $a_y$ and $a_z$ explicitly.

\begin{align}
    a_y &= \frac{K_2}{\cos \theta} \left( 1 - a_x \right) \\
    &= \frac{K_2}{\cos \theta} \left( \frac{2}{\left( \frac{K_2}{\cos \theta} \right)^2 + \left( \frac{K_1}{\sin \theta} \right)^2 + 1} \right) \\
    a_z &= \frac{K_1}{\sin \theta} \left( 1 - a_x \right) \\
    &= \frac{K_1}{\sin \theta} \left( \frac{2}{\left( \frac{K_2}{\cos \theta} \right)^2 + \left( \frac{K_1}{\sin \theta} \right)^2 + 1} \right) \\
\end{align}

To simplify our presentation, we consider the angle of a scaled vector $\mathbf{f}$ that is collinear to $\hat{\mathbf{a}}$ to remove the denominator from its components.

\begin{align}
\mathbf{f} &= \left( \left( \frac{K_2}{\cos \theta} \right)^2 + \left( \frac{K_1}{\sin \theta} \right)^2 + 1 \right) \hat{\mathbf{a}} \\
f_x &= \left( \frac{K_2}{\cos \theta} \right)^2 + \left( \frac{K_1}{\sin \theta} \right)^2 - 1 \\
f_y &= \frac{2 K_2}{\cos \theta} \\
f_z &= \frac{2 K_1}{\sin \theta}
\end{align}

We then solve for the angular displacement $\phi$ using the $x-$ and $z-$components of $\mathbf{f}$ (Fig.~\ref{Figure3}).

\begin{align}
\phi &= \textrm{arctan2} \left( f_z, f_x \right) = \textrm{arctan2} \left( \frac{2K_1}{\sin \theta}, \left( \frac{K_2}{\cos \theta} \right)^2 + \left( \frac{K_1}{\sin \theta} \right)^2 - 1 \right) 
\end{align}

Similarly, we solve for the angular displacement $\psi$ using the $x-$ and $y-$components of $\mathbf{f}$ (Fig.~\ref{Figure4}).

\begin{align}
\psi &= \textrm{arctan} \left( \frac{f_y}{f_x} \right) = \left( \frac{2K_2}{\cos \theta\left[ \left( \frac{K_2}{\cos \theta} \right)^2 + \left( \frac{K_1}{\sin \theta} \right)^2 - 1 \right]} \right)
\end{align}

\section{Validation}

\begin{figure}[b]
	\includegraphics[width=160mm]{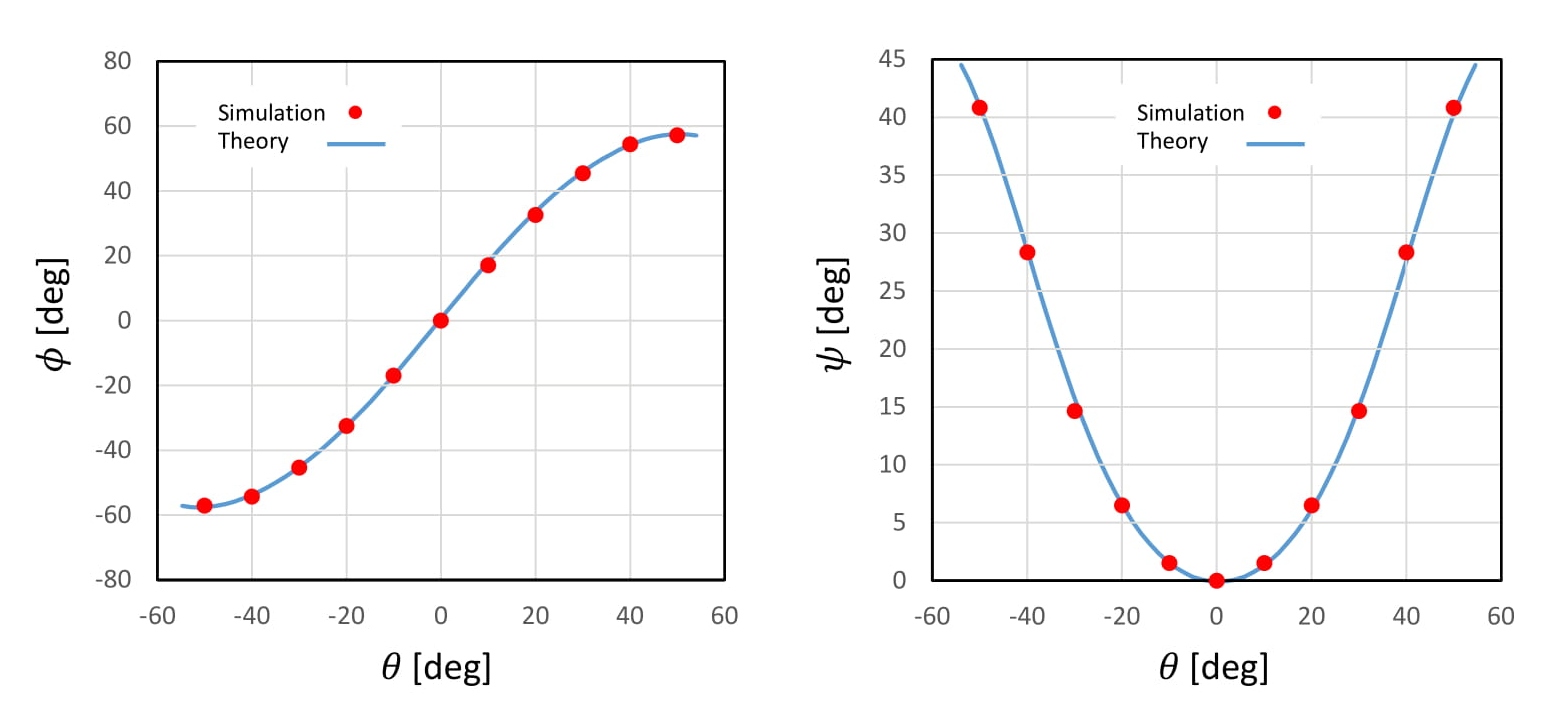}
	\centering
	\caption{Plots show $\phi$ and $\psi$ values as a function of $\theta$. The agreement between the analytical results and simulation validates the developed equations characterizing the folding of asymmetric Miura vertex.}
	\label{Figure5}
\end{figure}

In this section, we perform a kinematics simulation to validate the developed explicit analytical relationship between the fold pattern, fold angle, and displacement angles of the asymmetric Miura vertex (equations 38 and 39). Fig.~\ref{Figure5} shows the variation of $\phi$ and $\psi$ as a function of $\theta$ for a Miura vertex with $\alpha_1=30^{\circ}$ and $\alpha_2=60^{\circ}$. These specific $\alpha$ angle values have been randomly selected only for the purpose of validation. The blue solid lines in both plots demonstrate the results from analytical relationships and cover the $\theta$ axis from $-55^{\circ}$ to $55^{\circ}$. In asymmetric Miura vertex the limits of angle $\theta$, range of folding, can be calculated from the following equation as a function of $\alpha$ angles \cite{kamrava2018slender}:

\begin{align}
\theta _{ ~limit} &= \pm \left(90^{\circ}-\frac{1}{2} \cos^{-1}\left(\frac{\tan \alpha_1}{\tan \alpha_2}\right)\right)
\end{align}

Moreover, we used a commercially available software, SolidWorks (Dassault Systems, Vlizy-Villacoublay, France) to simulate the folding of an asymmetric Miura vertex with $\alpha_1=30^{\circ}$ and $\alpha_2=60^{\circ}$. The software solves the rigid-body equations of motion numerically and finds geometrically admissible configurations \cite{chang2019motion}. Using this technique we determine the relation between angle $\theta$ and angles $\phi$ and $\psi$. Red markers on the plots show the results extracted from the numerical simulation for 11 different folding configurations. The excellent agreement between the theoretical results (blue solid lines) and simulation (red markers) validates the derived explicit analytical folding relations in the asymmetric Miura vertex.

\section{Discussion}

In this paper we use a vector representation to derive an explicit analytical relationship between the fold pattern, fold angle, and displacement angles of the asymmetric Miura vertex. A numerical simulation technique was employed to validate the derived analytical relationships. Such an equation can be used to derive the requisite design parameters of an origami string based on its desired spatial trajectory. Future work could reveal a similar closed-form solution for the more general case of arbitrary four-crease vertices (ones without collinear spinal creases).
\bibliographystyle{IEEEtran}
\bibliography{template}

\end{document}